\documentclass{PoS}

\newcommand{\pom}{\mbox{\scriptsize\it I\hspace{-0.5ex}P}}
\newcommand{\Pom}{\mbox{\it I\hspace{-0.5ex}P}}

\title{Diffractive production of isolated photons with the ZEUS detector at HERA}

\ShortTitle{Diffractive isolated photons}

\author{\speaker{Aharon Levy}\\
     On behalf of the ZEUS Collaboration\\Tel Aviv University, Tel Aviv, Israel\\
      \email{levyaron@post.tau.ac.il}}

\abstract{ 
The photoproduction of isolated photons has been measured in 
diffractive events recorded by the ZEUS detector at HERA.  Cross
sections are evaluated in the photon transverse-energy and
pseudorapidity ranges
 5 $< E_T^{\gamma} < $15 GeV and $-0.7 < \eta^{\gamma} <
0.9$ inclusively and also with a jet with transverse energy and
pseudorapidity in the ranges 4$ < E_T^{jet} <$ 35 GeV and  -1.5$ <
\eta^{jet} <$ 1.8, using a total integrated electron--proton luminosity of 374
$\mathrm{pb}^{-1}$. A number of kinematic variables were studied and
compared to predictions from the \textsc{Rapgap} Monte Carlo model.  An
excess of data is observed above the \textsc{Rapgap} predictions for $z_{\pom }^{\mathrm {meas}} >
0.9$, where $z_{\pom }^{\mathrm {meas}}$ is the fraction of the longitudinal momentum of the
colourless ``Pomeron'' exchange that is transferred to the photon--jet
final state, giving evidence for direct Pomeron interactions.
}

\FullConference{EPS-HEP 2017, European Physical Society conference on High Energy Physics\\
		5-12 July 2017\\
		Venice, Italy}

\begin{document}

\section{Introduction}

The HERA $ep$ collider took data during the period 1992 - 2007. Its rich physics output covered hard processes in the deep inelastic scattering (DIS) regime down to soft interactions in the photoproduction region~\cite{HAAC}. Hard processes were studied also in the photoproduction regime when a high scale was present. In the present talk, results are presented of hard scattered isolated photons associated with jets in the photoproduction region. The photoptroduction regime is usually defined as a region where the four-momentum squared between the scatted and the incoming electron, $Q^2$, is less than 1 GeV$^2$. In this region no scattered electron is observed.

HERA also showed that diffraction, usually thought to be a soft process, is present also in hard processes~\cite{LRG} resulting in events with a large rapidity gap. The results presented in this talk are of isolated photons (called hereafter prompt photons)  which are diffractively produced. Analyses of prompt photons in the DIS~\cite{DISg} and the non-diffractive photoproduction~\cite{PHPg} regions have already been presented by the two HERA collaborations.

The reaction studied here can be written as
\begin{equation}
e^\pm + p \to (e^\pm) + \gamma + X + {\rm LRG} + ({\rm p \ \ or \ \  pdis}) ,
\label{Eq:reaction}
\end {equation}
where LRG is a large rapidity gap and pdis stands for proton dissociation. The LRG is also expressed by a the variable $\eta_{max}$, which is the maximum pseudorapidity for energy-flow objects with energy of 0.4~GeV. Typially for diffraction events $\eta_{max} < $2.5.
The brackets on the right-hand side of the reaction describe states that are not observed in the detector. The prompt photon has a high transverse energy, $E_T >$ 5 GeV, and $X$ represents hadrons or jets.
The reaction~(\ref{Eq:reaction}) can be described as a reaction between the virtual photon, $\gamma*$ at thte electron vertex and the Pomeron, \Pom,  at the proton vertex. The fraction of thte proton energy carried by the Pomeron is given to a good approximation by 
\begin{equation}
x_{\pom} =(E^{\rm {all}}+ p_Z^{\rm {all}})/2E_p,
\end{equation} 
where $E_p$ is the energy of the proton beam.  

Having a hard scale, the partonic structure of their interaction can be studied. Both the exchanged photon and the Pomeron can act either as a whole (direct) or as a source of partons (resolved). These two classes of process, which
are unambiguously defined only at the leading order (LO) of QCD, may be partially
distinguished in events containing a high-$E_T$ photon and a jet by
means of the quantity
\begin{equation}
x_{\gamma}^{\mathrm{meas}} =  \frac{E^{\gamma}+E^{\rm jet}-p_Z^\gamma-p_Z^{\rm jet}}
{E^{\rm {all}}- p_Z^{\rm {all}}}, 
\end{equation}
which measures the fraction of the incoming photon energy that is
given to the outgoing photon and jet.  The quantities $E^{\gamma}$ and
$E^{\rm jet}$ denote the energies of the outgoing photon and the jet,
respectively, and $p_Z$ denotes the corresponding longitudinal
momenta. The suffix ``all'' refers to all objects that are measured in
the detector or, in the case of simulations at the hadron level, all
final-state particles except for the scattered beam electron and the
outgoing proton. Events with a detected final-state electron are
excluded from this analysis.

The Pomeron may be described analogously to the
photon~\cite{zp:c65:657,mrw}. The fraction of the Pomeron energy that
takes part in the hard interaction that generates the outgoing photon
and jet is given by~\cite{epj:c70:15}:
\begin{equation} 
z_{\pom }^{\mathrm {meas}} =  \frac{E^{\gamma}+E^{\rm jet}+p_Z^\gamma+p_Z^{\rm jet}}{E^{\rm {all}}+ p_Z^{\rm {all}}},
\end{equation}
where the quantities are as before, and $z_{\pom}=1$ corresponds to
direct Pomeron events, which are equivalent to the presence of a delta-function
in the parton distribution functions at $z_{\pom}=1$~\cite{zp:c65:657,mrw}.
An event whose observed final
state consists only of a prompt photon and a jet has $x^\gamma=z_{\pom}=1.$

\begin{figure}[h!]
\hspace*{0.02\linewidth}\includegraphics[width=0.6\linewidth]{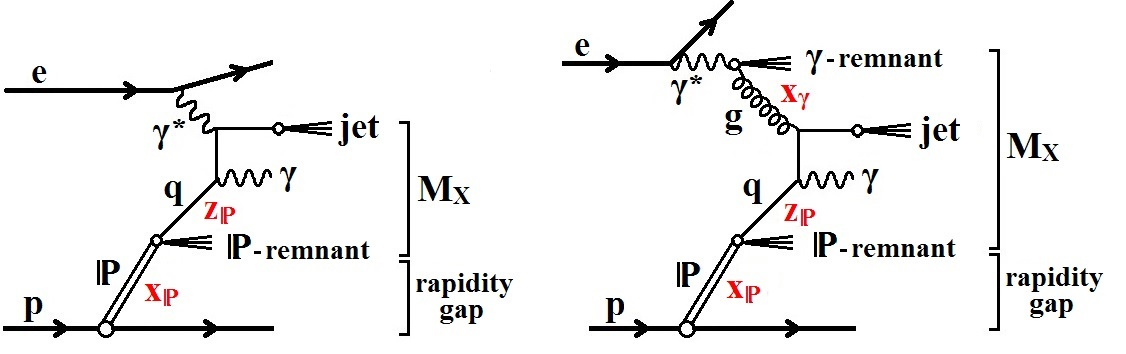}
\hspace*{0.02\linewidth}\includegraphics[width=0.35\linewidth]{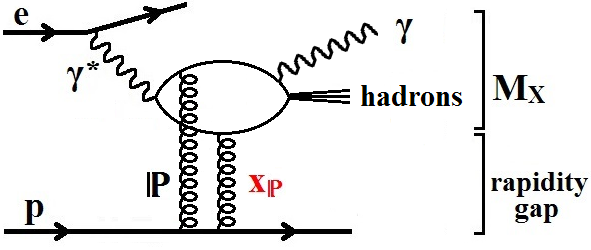}
~\\[1mm] \hspace*{1.5cm} (a) \hspace{4cm} (b) \hspace{4.5cm} (c)
\vspace*{0mm}
\caption{\small
Examples of diagrams for the diffractive 
production of a prompt photon and a jet in $ep$ scattering from (a)
direct (b) resolved photons, interacting with a resolved Pomeron. 
(c) Example of an interaction
between a direct photon and a direct Pomeron.
}
\label{fig1}
\end{figure}

Examples of diagrams describing the above processes are shown in Fig.~\ref{fig1}. A reaction which proceeds through the interaction of a direct photon with a quark from a resolved Pomeron is shown in Fig.~\ref{fig1}(a). In (b) a gluon from a resolved photon interacts with a quark from a resolved Pomeron. The interaction of a direct photon with a direct Pomeron is shown in Fig.~\ref{fig1}(c).

The data used in this analysis is based on a data samples corresponding
to an integrated luminosity of $374\,\mathrm{pb}^{-1}$, taken
during the years 2004--2007  with the
ZEUS detector at HERA. During this period, HERA ran with electron and positron
beams\footnote{Hereafter, ``electron'' refers to both electrons and
positrons.} of energy $E_e = 27.5$ GeV and a proton beam of energy
$E_p =920$ GeV.

\section{The data}

The following selections were made:
\begin{itemize}
\item
the forward scattered proton was not measured;
\item
non-diffractive events were removed by the cuts $\eta_{max} < $2.5 and  $x_{\pom} <$ 0.03;
\item
remaining DIS events were removed by excluding events with identified electrons;
\item
Bethe-Heitler ($ep\to ep\gamma$) and deeply virtual Compton events were removed;
\item
requirement was made for hard isolated photon candidates ($E_T^\gamma > $5~GeV) in the barrel calorimeter ($-0.7 < \eta^\gamma < 0.9$);
\item
high-p$_T$ jets, using a $k_T$ algorithm, were required to be in the region $-1.5 < \eta^{jet} < 1.8$.
\end{itemize}

\section{The outgoing  prompt photon}

The hard scattered photons can be extracted in a region of the detector which is finely segmented in the $Z$ direction, namely the electromagnetic section of the ZEUS barrel calorimeter (BEMC), limiting the pseudorapidity of the photon to the region 
$-0.7<\eta^\gamma<0.9$. This fine granularity allows the use of shower-shape distributions to distinguish isolated photons from the product of neutral meson decays such as $\pi^0\to\gamma\gamma$. To this end, a method was 
developed~\cite{DISg} to make use of the energy-weighted width, measured in the $Z$ direction, of the BEMC energy cluster comprising the photon candidate,
\begin{equation}
\langle\delta Z\rangle=
\sum \limits_i E_i|Z_i-Z_{\mathrm{cluster}}|
\mathrm{\,\left/\,\right.}( w_{\mathrm{cell}}\sum \limits_i E_i),
\end{equation}
where $Z_{i}$ is
the $Z$ position of the centre of the $i$-th cell,
$Z_{\mathrm{cluster}}$ is the energy-weighted centroid of the energy-flow-object
cluster, $w_{\mathrm{cell}}$ is the width of the cell in the $Z$
direction, and $E_i$ is the energy recorded in the cell. The sum runs
over all BEMC cells in the cluster.  

\begin{figure}[h!]
  \centering
   \includegraphics[width=0.45\textwidth]{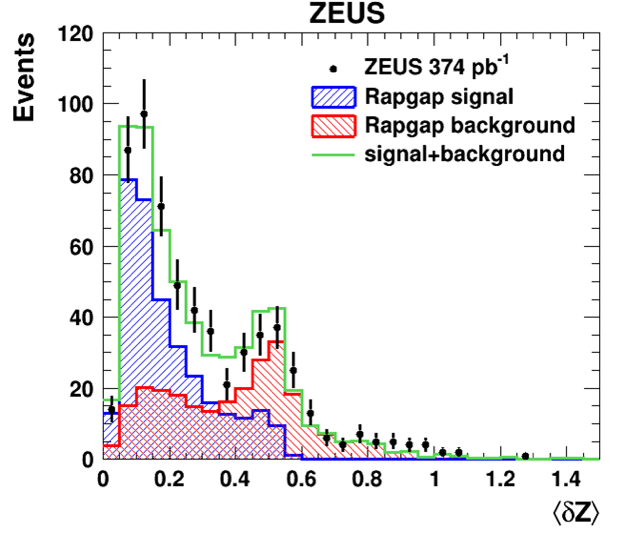}
\vspace*{0mm}
\caption{\small 
Distribution of $\langle \delta Z \rangle$ for selected diffractive
events with a photon candidate and at least one jet. The error bars denote the statistical uncertainties on the data,
which are compared to the fitted signal and background components from
the MC.  The unit of measurement of $\langle \delta Z \rangle$ is the
width of one BEMC cell.  }
\label{fig:showers}
\end{figure}

As this analysis studies diffractive prompt photon events having a large rapidity gap, the \textsc{Rapgap} Monte Carlo generator has been use to simulate the expectted $\delta Z$ distribution of prompt photons coming from a signal event and the distribution of photons coming from the background. In each bin of each physical quantity, the data were fitted to a combination of photon signal and hadronic background. The data together with the fitted results, for events with photon candidastes and at least one jet, are shown in Fig.~\ref{fig:showers}. A good fit to thte data is obtained allowing to extract statistically the photon signal. 

\section{Results}

\begin{figure}[h!]
\begin{center}
\includegraphics[width=0.45\linewidth]{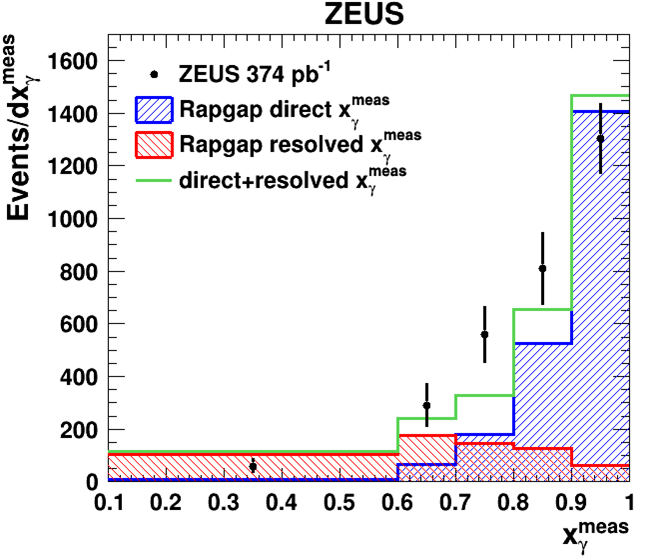}  
\end{center}
\vspace*{0mm}
\caption{ \small 
Events with a photon and at least one jet as a function of $x_{\gamma}^{\mathrm{meas}}$, per
unit interval in $x_{\gamma}^{\mathrm{meas}}$, compared to a normalised 70:30 mixture of
direct:resolved photon \textsc{Rapgap}  events without reweighting.}
\label{fig:xgam}
\end{figure}

After selecting the photon candidate events as described above, the distribution of the 
$x_{\gamma}^{\mathrm{meas}}$ is shown in Fig.~\ref{fig:xgam}. It peaks toward high $x_{\gamma}^{\mathrm{meas}}$ and a mixture of 70:30 of direct and resolved photon component events, as obtained from \textsc{Rapgap} gives a reasonable description of the data.

\begin{figure}[h!]
\begin{center}
\includegraphics[width=0.45\linewidth]{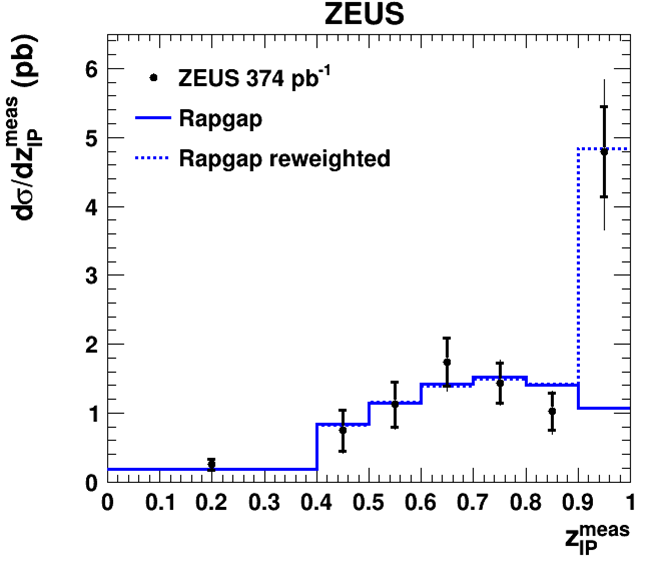}
\end{center}
\caption{\small 
Differential cross section for isolated photon production accompanied
by at least one jet, as a function of $z_{\pom }^{\mathrm {meas}}$.
The unreweighted \textsc{Rapgap} prediction is normalised to the data integrated over 
the region $z_{\pom }^{\mathrm {meas}}< 0.9$; the reweighted prediction is normalised
to the full integrated data.
}
\label{fig:xszpom}
\end{figure}

The distribution of $z_{\pom }^{\mathrm {meas}}$ is shown in Fig.~\ref{fig:xszpom}. A sharp peak is observed for $z_{\pom }^{\mathrm {meas}} >$ 0.9 in contrast to the prediction of the MC generator that described well the $x_{\gamma}^{\mathrm{meas}}$. The \textsc{Rapgap} generator does not include the process of a direct Pomeron, as presented in Fig.~\ref{fig1}(c). Agreement with the data is obtained by reweighting \textsc{rapgap} (a weight of 7 is applied to the direct photon component). The reweighted \textsc{Rapgap} changes very little in the description of Fig.~\ref{fig:xgam} and still provides an adequate description of the data.

The excess observed for $z_{\pom }^{\mathrm {meas}} >$ 0.9 is evidence for the presence of a direct Pomeron interaction, predominantly in e direct photon channel.

%\section*{Acknowledgements}

\end{document}